\documentclass[aps,prd,floatfix,superscriptaddress,showpacs,showkeys]{revtex4}
\usepackage{graphicx,epsfig,epstopdf}
\usepackage{amssymb,amsmath,amsxtra,amsfonts}
\usepackage{bm}

\usepackage{longtable}
\usepackage{multirow}
\usepackage{booktabs}
\usepackage{array}
\usepackage{wrapfig}
\usepackage{color}

\usepackage{titlesec}
\titleformat{\section}{\large\bfseries}{\thesection}{1em}{}

\newcommand{\bea}{\begin{eqnarray}}
\newcommand{\ena}{\end{eqnarray}}
\newcommand{\be}{\begin{equation}}
\newcommand{\en}{\end{equation}}
\newcommand{\nn}{\nonumber\\}
\newcommand{\ed}{\end{document}} 
\newcommand{\Tr}{\mbox{\rm{tr}}}

\begin{document}

\hfill MITP/17-043 (Mainz)

\title{$Z_b(10610)$ and $Z_b'(10650)$ decays 
in a covariant quark model} 

\author{Fabian~Goerke}
\affiliation{Institut f\"ur Theoretische Physik, Universit\"at T\"ubingen,
Kepler Center for Astro and Particle Physics,
Auf der Morgenstelle 14, D-72076, T\"ubingen, Germany}

\author{Thomas~Gutsche}
\affiliation{Institut f\"ur Theoretische Physik, Universit\"at T\"ubingen,
Kepler Center for Astro and Particle Physics,
Auf der Morgenstelle 14, D-72076, T\"ubingen, Germany}

\author{Mikhail~A.~Ivanov}
\affiliation{Bogoliubov Laboratory of Theoretical Physics, 
Joint Institute for Nuclear Research, Dubna, Russia}

\author{J\"urgen~G.~K\"orner}
\affiliation{PRISMA Cluster of Excellence, Institut f\"{u}r Physik, 
Johannes Gutenberg-Universit\"{a}t,  
D-55099 Mainz, Germany}

\author{Valery~E.~Lyubovitskij}
\affiliation{Institut f\"ur Theoretische Physik, Universit\"at T\"ubingen,
Kepler Center for Astro and Particle Physics,
Auf der Morgenstelle 14, D-72076, T\"ubingen, Germany}
\affiliation{Departamento de F\'\i sica y Centro Cient\'\i fico
Tecnol\'ogico de Valpara\'\i so-CCTVal, Universidad T\'ecnica
Federico Santa Mar\'\i a, Casilla 110-V, Valpara\'\i so, Chile}
\affiliation{Department of Physics, Tomsk State University,  
634050 Tomsk, Russia} 
\affiliation{Laboratory of Particle Physics, Tomsk Polytechnic University, 
634050 Tomsk, Russia} 

\begin{abstract}
We present a calculation of the strong decays of the   
exotic states $Z_b(10610)$ and $Z_b'(10650)$ using  
a covariant quark model. We use a molecular-type four-quark 
current for the coupling of the $Z_b(10610)$ and $Z_b'(10650)$
to the constituent heavy and light quarks.
\pacs{13.20.Gd,13.25.Gv,14.40.Rt,14.65.Fy}

\keywords{quark model, confinement, 
exotic states, tetraquarks, decay widths}

\end{abstract}

\maketitle

\section{Introduction}
\label{sec:intro}

A few years ago the Belle Collaboration~\cite{Belle:2011aa}
reported on the observation of two charged bottomoniumlike resonances
in the mass spectra of $\pi^{\pm}\Upsilon(nS)$ ($n=1,2,3$) and
$\pi^{\pm}h_b(mP)$ ($m=1,2$) in the decays 
$\Upsilon(5S)\to\Upsilon(nS)\pi^+\pi^-, h_b(mP)\pi^+\pi^-$. 
The measured masses and widths were given by
\bea 
      M_{Z_b} &=& (10607.2 \pm 2.0 ) \ \text{MeV}\,, \qquad  
  \Gamma_{Z_b} = ( 18.4 \pm 2.4 ) \ \mathrm{MeV}\,,
\nn
M_{Z'_b} &=& ( 10652.2 \pm 1.5) \ \mathrm{MeV}\,, \qquad 
\Gamma_{Z'_b} = ( 11.5 \pm 2.2 ) \ \mathrm{MeV}\,.
\label{eq:Belle-1} 
\ena
The existence of these two states was later confirmed by the same
collaboration~\cite{Garmash:2014dhx,Garmash:2015rfd} in differing
decay channels.

In~\cite{Garmash:2014dhx} the Belle Collaboration rediscovered the
two states in $e^+e^-$
annihilation into $\Upsilon(nS) \pi^+\pi^-$ ($n=1,2,3$).
It was found that the favored quantum numbers for both $Z_b$ states are 
$I^G(J^P)=1^+(1^+)$.

In the paper~\cite{Garmash:2015rfd} the Belle Collaboration reported on
the results of an analysis of the three-body processes
$e^+e^-\to B\bar B\pi^\pm, B\bar B^\ast\pi^\pm$, and 
$ B^\ast\bar B^\ast\pi^\pm$.
It was found that the transitions $Z^\pm_b(10610)\to [B\bar B^\ast + c.c.]^\pm$
and  
$Z^\pm_b(10650)\to [B^\ast\bar B^\ast]^\pm$ dominate among
the corresponding final states. The fit to the $BB^\ast\pi$ and
$B^\ast B^\ast\pi$ data gave the values of $Z_b$ masses and widths 
\bea 
      M_{Z_b} &=& (10605 \pm 6 ) \ \text{MeV}\,, \qquad  
  \Gamma_{Z_b} = ( 25 \pm 7 ) \ \mathrm{MeV}\,,
\nn
M_{Z'_b} &=& ( 10648 \pm 13) \ \mathrm{MeV}\,, \qquad 
\Gamma_{Z'_b} = ( 23 \pm 8 ) \ \mathrm{MeV}\,.
\label{eq:Belle-2} 
\ena
The relative decay fractions for the $Z_b$-decays
were determined assuming that they are saturated by 
the $\Upsilon(nS)\pi$ ($n=1,2,3$), $h_b(mP)\pi$ ($m=1,2$) 
and $B^{(\ast)}\bar B^\ast$ channels.

The theoretical structure assignments for the two hidden-bottom
meson resonances proposed immediately after their 
observation~\cite{Bondar:2011ev}-\cite{Gutsche:2017oro} 
were based on a molecular~\cite{Bondar:2011ev,Voloshin:2011qa,%
Mqh:2011,Pjl:2011,Slzhu:2011,Cleven:2011gp,Dong:2012hc}  
and a tetra-quark interpretation~\cite{Guo:2011gu,Richard:2011}
using the analogy to the corresponding states in the charm sector.
In~\cite{Danilkin:2011sh}
the new resonances were identified as a hadro-quarkonium system
based on the coupling of light and heavy quarkonia to
intermediate open-flavor heavy-light mesons. Subsequently these states 
have been studied extensively using various assignments within 
different approaches: 
chiral quark model~\cite{Li:2012wf}, using phenomenological 
Lagrangians~\cite{Chen:2012yr,Li:2012as,Ohkoda:2013cea}, 
effective field theory~\cite{Cleven:2013sq,Huo:2015uka}, 
QCD sum rules~\cite{Wang:2013zra}, 
meson exchange model~\cite{Dias:2014pva}, 
effective range theory~\cite{Kang:2016ezb}, 
and holographic QCD~\cite{Gutsche:2017oro}. 

In Ref.~\cite{Dong:2012hc} some of us presented a 
detailed analysis of the strong decays  
$Z_b^+ \to \Upsilon(nS) + \pi^+$ and 
$Z_b^{\prime +} \to \Upsilon(nS) + \pi^+$, $n=1,2,3$) using a phenomenological
Lagrangians formulated in 
terms of hadronic degrees of freedom. The hadronic molecular 
approach was 
developed in Refs.~\cite{hmappr} and is based on the compositeness
condition formulated in~\cite{Weinberg:1962hj,Efimov:1993ei}. 
The compositeness condition implies that the renormalization
constant of the hadron wave function is set equal to zero or that the hadron
is a bound state of its constituents.

In this paper we study the strong decays of the $Z_b$ and $Z_b'$ states using 
the covariant confined quark model (CCQM) proposed in 
Refs.~\cite{Dubnicka:2010kz,Dubnicka:2011mm}. The CCQM has been successfully 
applied to the description of the properties of the exotic 
$X(3872)$, $Z_c(3900)$, $Z(4430)$, and $X(5568)$ 
states~\cite{Dubnicka:2010kz,Dubnicka:2011mm,% 
Gutsche:2016cml,Goerke:2016hxf}. 
The present study complements the analysis performed 
in Ref.~\cite{Dong:2012hc}. Both approaches are based on the use 
of the compositeness condition once formulated in terms of hadronic
constituents and then in terms of quark constituents. 
The four-quark approach gives an opportunity 
to analyze both tetraquark and molecular configurations of constituents 
inside exotic states. In Ref.~\cite{Goerke:2016hxf} we have shown that 
the four-quark picture with molecular-type of interpolating currents are
favored for the $Z_c(3900)$ states. In the case of the $Z_b(10610)$ and 
$Z_b'(10650)$ states there is strong experimental confirmation that they 
are four-quark states with a molecular-type configuration. For this reason 
we will only consider molecular-type four-quark currents for these states
in the present paper. 

The paper is organized as follows. In Sec.~\ref{sec:Zb-states}, we 
discuss the choice of the interpolating four-quark currents with 
molecular configuration for the exotic $Z_b(10610)$ and $Z_b'(10650)$ states.
We give model independent formulas for the matrix elements and decay rates
of the strong two-body decays of these states. 
In Sec.~\ref{sec:model-calculas} we calculate matrix elements
and decays rates in the framework of our covariant quark model.
In Sec.~\ref{sec:numerics} we discuss the numerical results obtained
in our approach and compare them with available experimental data.
Finally, in  Sec.~\ref{sec:Summary} we summarize our findings.

%--------------------------------------------------------------------

\section{$Z_b(10610)$ and $Z_b(10650)$ 
as molecular-type four-quark states} 
\label{sec:Zb-states}
Since the masses of the $Z^+_b(10610)$ and $Z_b^\prime(10650)$ resonances
are very close to the respective $B^\ast\bar B$ (10604 MeV)
and  $B^\ast\bar B^\ast$ (10649 MeV) thresholds, 
in Ref.~\cite{Bondar:2011ev} it was suggested  
that they have molecular-type binding structures. 
Their observed quantum numbers are $I^G(J^P)=1^+(1^+)$, so that the neutral
isotopic states with $I_3=0$ have the quantum numbers $J^{PC}=1^{+-}$. 
As a result the allowed interpolating four-quark currents have the form:  
\bea
J^\mu_{Z_b^+} &=& \frac{1}{\sqrt{2}} 
\left[ (\bar d \gamma_5 b) (\bar b  \gamma^\mu u)
      +(\bar d \gamma^\mu b)(\bar b \gamma_5  u) \right]\,, 
\label{eq:Zb-cur}\\[2mm]
J^{\mu\nu}_{Z_b^{\prime +}} &=& \varepsilon^{\mu\nu\alpha\beta} 
(\bar d \gamma_\alpha b) (\bar b \gamma_\beta u)
\label{eq:Zb1-cur} \,. 
\ena 
Such a choice guarantees that the $Z_b$-state can only decay to the 
$[\bar B^\ast B+c.c.]$ pair whereas  the $Z'_b$-state can decay only to 
a $\bar B^\ast B^\ast$ pair. Decays into the $BB$-channels are forbidden.  

The exotic states can also decay into a bottomonium state plus
a charged light meson. We start with the classification of such
two-body decays. In Table~\ref{tab:bb} we show a list of the orbital 
excitations of the $b\bar b$ states with spin 0 and 1 by analogy with 
the charmonium excitations as given in Ref.~ \cite{Ivanov:2005fd}.
\begin{table}[ht]
\begin{center}
\caption{The  bottomonium  states $^{2S+1}L_{\,J}$.
We use the notation $\stackrel{\leftrightarrow}{\partial}=
\stackrel{\rightarrow}{\partial}-\stackrel{\leftarrow}{\partial}$.}
\label{tab:bb}
\def\arraystretch{1.5}
\begin{tabular}{|c|c|c|c|}
\hline
quantum number $I^G(J^{PC})$ & name & quark current & mass (MeV) \\
\hline\hline
 $0^+(0^{-+})$ ($S=0, L=0$) & $^1S_0=\eta_b(1S)$ & $\bar b\, i\gamma^5\, b $ &
$ 9399.00\pm 2.30$ \\
\hline
 $0^-(1^{--})$ ($S=1, L=0$) & $^3S_1=\Upsilon$ & $\bar b\,\gamma^\mu\, b $ &
$ 9460.30\pm 0.26$ \\
\hline
 $0^+(0^{++})$ ($S=1, L=1$) & $^3P_0=\chi_{b0}$ & $\bar b\, b $ &
 $ 9859.44\pm 0.52 $\\
\hline
 $0^+(1^{++})$ ($S=1, L=1$) & $^3P_1=\chi_{b1}$ & $\bar b\, \gamma^\mu\gamma^5\, b $
 & $ 9892.72\pm 0.40$ \\
\hline
$0^-(1^{+-})$ ($S=0, L=1$) & $^1P_1=h_b(1P)$ &
 $\bar b\, \stackrel{\leftrightarrow}{\partial}^{\,\mu} \gamma^5\, b $
& $ 9899.30\pm 0.80$ \\
\hline
\end{tabular}
\end{center}
\end{table}

$G$-parity is a multiplicatively quantum number conserved 
in strong interactions. Keeping in mind that $G(\pi^+)=-1$ and 
$G(\rho^+)=+1$, the decays $Z_b\to \Upsilon\rho,\eta_b\pi,\chi_{b1}\pi,h_b\rho$
are forbidden. The decay $Z_b\to \chi_{b1}\rho$ is not allowed kinematically.
There are therefore only the three allowed decays: 
$Z^+_b\to \Upsilon+\pi^+$, $Z^+_b\to h_b+\pi^+$ and  $Z^+_b\to \eta_b+\rho^+$.
 
Let us discuss the spin kinematics for the three decays
$1^+\to 1^- + 0^-$ ($Z^+_b\to \Upsilon+\pi^+$, 
                   $Z_b^+\to [\bar B^{\ast\,0} B^+ + c.c.]$, 
                   $Z^+_b\to \eta_b+\rho^+$),
$1^+\to 1^+ + 0^-$ ( $Z^+_b\to h_b+\pi^+$) 
and
$1^+\to 1^- + 1^-$ ( $Z_b^+\to \bar B^{\ast\,0} B^{\ast\,+}$). 
 
\begin{itemize}
\item The decay $1^+\to 1^- + 0^-$.
\end{itemize}

The momenta and the Lorentz indices of the polarization four-vectors
in the decay are labelled according to the transition matrix element
\be
M=\langle 1^- (q_1;\delta), 0^-(q_2) |\,T\,|1^+(p;\mu)\rangle \,.
\label{eq:1m0m}
\en
The product of the parities of the two final state mesons 
is $(+1)$ which matches the parity of the initial state. 
Thus the two final state mesons must have even relative orbital momenta. 
In the present case these are $L=0,2$. 
The spins $s_1$ and $s_2$ of the two final state mesons couple to the total 
spin $S=1$. Thus one has the two $(LS)$ amplitudes $(L=0,S=1)$
and  $(L=2,S=1)$. The covariant expansion of the transition matrix is given by
\be
M\,=\,(A\,g^{\mu \delta}+ B\,q_1^\mu q_2^\delta)\,
\varepsilon_\mu\,\varepsilon^\ast_{1\delta}\,.
\label{eq:A3}
\en
Alternatively one may describe the transition amplitude by the
helicity amplitudes $H_{\lambda\lambda_1}$. The helicity amplitudes may be
expressed as a linear
superposition of the invariant amplitudes $A$ and $B$. One has
\be
H_{00} = -\,\frac{E_1}{M_1}\,A - \frac{M}{M_1}{|\bf q_1}|^2 \,B\,, \qquad
H_{+1+1} = H_{-1-1}=-\,A.
\label{eq:A5}
\en
Since the particles of the 
initial and final states are on their mass-shells one
has $p^2=M^2$, $q_1^2=M_1^2$, $q_2^2=M_2^2$ and $p^\mu\varepsilon_\mu=0$,
$q_1^\delta\varepsilon^\ast_\delta=0$. The magnitudes of the final state 
three-momentum and energy in the rest frame of the initial particle 
is given by $|{\bf q_1}|=\lambda^{1/2}(M^2,M_1^2,M_2^2)/2M$
and $E_1=(M^2+M_1^2-M_2^2)/2M$, respectively.

The rate of the decay $1^+(p)\to 1^-(q_1) + 0^-(q_2)$ finally reads
\bea
\Gamma &=&
\frac{\bf|q_1|}{24\pi M^2}
\Big\{ 
\Big( 3 + \frac{ {\bf|q_1|}^2}{M^2_1}\Big)\,A^2
+ (M^2+M_1^2-M_2^2)\,\frac{ {\bf|q_1|}^2}{M_1^2}\, A\,B
+ \frac{M^2}{M_1^2}{\bf|q_1|}^4\,B^2 \Big\}\,,
\nn[2ex]
&= & \frac{\bf|q_1|}{24\pi M^2}
\Big\{ |H_{+1+1}|^2 +  |H_{-1-1}|^2 +  |H_{00}|^2 \Big\}\,.
\label{eq:width1m0m}
\ena
 
\begin{itemize}
\item The decay $1^+\to 1^+ + 0^-$.
\end{itemize}

The matrix element is described by 
\be
M=\langle 1^+ (q_1;\delta), 0^-(q_2) |\,T\,|1^+(p;\mu)\rangle \,.
\label{eq:1p0m}
\en
In this decay the product of the parities of the two final state mesons 
is $(-1)$ which does not match the parity of the initial state. 
Thus the two final state mesons must have odd relative orbital momenta. 
In the present case this is $L=1$. 
The spins $s_1$ and $s_2$ of the two final state mesons couple to the total 
spin $S=1$. Thus one has only one $(LS)$ amplitude with $L=1, S=1$.
This implies that there is only one invariant amplitude in this transition.
Accordingly there is only one term in the covariant expansion of the matrix
given by
\be
M\,=\,C\,\varepsilon^{q_1q_2\mu\delta}\,
\varepsilon_\mu\,\varepsilon^\ast_{1\delta}\,,
\label{eq:M1p0m}
\en 
where 
$\varepsilon^{q_1q_2\mu\delta} = 
q_{1\alpha} q_{2_\beta} \, \varepsilon^{\alpha\beta\mu\delta}$. 

The rate of the decay $1^+(p)\to 1^+(q_1) + 0^-(q_2)$ can be seen to be given
by
\bea
\Gamma &=&
\frac{\bf|q_1|^3}{12\pi M^2}\,C^2\,.
\label{eq:width1p0m}
\ena
This decay is suppressed kinematically due to the
p-wave suppression factor ${\bf|q_1|^2}$.

\begin{itemize}
\item The decay $1^+\to 1^- + 1^-$.
\end{itemize}

By naive counting the covariant expansion of the matrix element 
\be
M=\langle 1^- (q_1;\delta), 1^-(q_2;\rho) |\,T\,|1^+(p;\mu)\rangle \,.
\label{eq:1m1m}
\en
involves five invariant amplitudes whereas there are only three independent
$(LS)$ amplitudes. This can be seen as follows.
The product of the parities of the two final state mesons 
is $(+1)$ which matches the parity of the initial state. 
Thus the two final state mesons must have even relative orbital momenta. 
In the present case these are $L=0,2$. 
The spins $s_1$ and $s_2$ of the two final state mesons couple to the total 
spins $S=0,1,2$. Thus one has three $(LS)$ amplitudes with $L=0, S=1$,
 $L=2, S=1$ and  $L=2, S=2$. 
Returning to the covariant form , the naive expansion of the matrix
element reads 
\be
M = (  A_1\, \varepsilon^{q_1\mu\rho\delta} 
     + A_2\, \varepsilon^{q_2\mu\rho\delta}
     + A_3\, \varepsilon^{q_1q_2\mu\rho}\,q_2^\delta
     + A_4\, \varepsilon^{q_1q_2\mu\delta}\,q_1^\rho
     + A_5\,\varepsilon^{q_1q_2\rho\delta}\,q_1^\mu )
     \varepsilon_\mu  \varepsilon^\ast_\delta  \varepsilon^\ast_\rho\,.  
\label{eq:covariants-1m1m}
\en
taking into account the transversality
conditions $p^\mu\varepsilon_\mu=0$, $q_1^\delta\varepsilon_\delta^*=0$ and
 $q_2^\rho\varepsilon_\rho^*=0$.
However, in four dimensions there are two linear relations between the five
covariants which can be calculated using the Schouten identity. In fact, one has
\bea
\varepsilon^{q_1q_2\mu\delta}\,q_1^\rho &=&
\varepsilon^{q_1q_2\rho\delta}\,q_1^\mu -
\varepsilon^{q_1\mu\rho\delta}\,q_1q_2 +
\varepsilon^{q_2\mu\rho\delta}\,q_1^2\,,
\nn[2mm]
\varepsilon^{q_1q_2\mu\rho}\,q_2^\delta\, &=&
\varepsilon^{q_1q_2\rho\delta}\,q_1^\mu +
\varepsilon^{q_1\mu\rho\delta}\,q^2_2  -
\varepsilon^{q_2\mu\rho\delta}\,q_1q_2\,.
\label{eq:Schouten}
\ena
The matrix element can therefore be written as
\be
M\,=\,\Big(  B_1\,\varepsilon^{q_1q_2\rho\delta}\,q_1^\mu 
           + B_2\,\varepsilon^{q_1\mu\rho\delta}
           + B_3\,\varepsilon^{q_2\mu\rho\delta} \Big)\,
\varepsilon_\mu\varepsilon_\delta\varepsilon_\rho\,.
\label{eq:1m1m-2}
\en 
where
\bea
B_1 &=& -( A_3 + A_4 + A_5)\,,
\nn
B_2 &=& -( A_1 + q_2^2 A_3 - q_1q_2 A_4) \,,
\nn
B_3 &=& -( A_2 - q_1q_2 A_3 + q_1^2 A_4)\,.
\label{eq:B}
\ena

The relation between the helicity amplitudes $H_{\lambda;\lambda_1\lambda_2}$
($\lambda=\lambda_1-\lambda_2$) and the invariant
amplitudes can be calculated to be
\bea
H_{\,0;+1+1} &=&  - H_{0;-1-1}
= -\,E_1\,A_1 -\,E_2\,A_2  - M{|\bf q_1}|^2 \,A_5\,, 
\nn
H_{+1;+1\,0} &=&  - H_{-1;-1\,0}
= \frac{(E_1M - M_1^2)}{M_2}\,A_1 +\,M_2\,A_2  
 - \frac{M^2}{M_2}{|\bf q_1}|^2 \,A_4\,, 
\nn
H_{-1;\,0+1} &=&  - H_{+1;\,0-1\,}
= M_1\,A_1 + \frac{(E_1M - M_1^2)}{M_1}\,A_2   
 - \frac{M^2}{M_1}{|\bf q_1}|^2 \,A_3\,. 
\label{eq:hel}
\ena

 The rate of the decay $1^+(p)\to 1^-(q_1) + 1^-(q_2)$, finally, reads
\bea
\Gamma &=&
\frac{\bf|q_1|}{24\pi M^2}\,
\cdot 2\,\Big\{
 M^2 {|\bf q_1}|^4 B_1^2 
+ \Big[ (1+\frac{M^2}{M_2^2}) {|\bf q_1}|^2 + 3 M_1^2\Big] B_2^2
+ \Big[ (1+\frac{M^2}{M_1^2}) {|\bf q_1}|^2 + 3 M_2^2\Big] B_3^2
\nn
& + & (M^2 + M_1^2 - M_2^2) {|\bf q_1}|^2 B_1B_2
+ (M^2 - M_1^2 + M_2^2) {|\bf q_1}|^2 B_1B_3
+ \Big[ 3 (M^2 - M_1^2 - M_2^2) -2 {|\bf q_1}|^2 \Big] B_2B_3
\Big\}
\nn
&=&
\frac{\bf|q_1|}{24\pi M^2}\,
\cdot 2\,\Big\{ |H_{\,0;+1+1}|^2 + |H_{+1;+1\,}|^2 + |H_{-1;\,0+1\,}|^2  \Big\}\,.
\label{eq:width1m1m}
\ena

\section{Two-body decays of $Z_b(10610)$ and $Z_b'(10650)$ 
in a covariant quark model} 
\label{sec:model-calculas}

The nonlocal renditions of the local four-quark currents written down in
Eqs.~(\ref{eq:Zb-cur}) and (\ref{eq:Zb1-cur}) are given by 
\bea
J^\mu_{Z_b^+}(x) &=& \int\! dx_1\ldots \int\! dx_4 
\delta\left(x-\sum\limits_{i=1}^4 w_i x_i\right) 
\Phi_{Z_b}\Big(\sum\limits_{i<j} (x_i-x_j)^2 \Big)
J^\mu_{4q}(x_1,\ldots,x_4),
\label{eq:nonlocal-mol-curZb}\\
J^\mu_{Z_b;4q}&=&  \frac{1}{\sqrt{2}} 
\Big\{
 (\bar d(x_3) \gamma_5   b(x_1))  (\bar b(x_2) \gamma^\mu u(x_4) )
+(\bar d(x_3) \gamma^\mu b(x_1))  (\bar b(x_2) \gamma_5  u(x_4) )
\,\Big\}
\nonumber
\ena 
and 
\bea
J^{\mu\nu}_{Z_b^{\prime +}}(x) &=& \int\! dx_1\ldots \int\! dx_4 
\delta\left(x-\sum\limits_{i=1}^4 w_i x_i\right) 
\Phi_{Z_b}\Big(\sum\limits_{i<j} (x_i-x_j)^2 \Big)
J^{\mu\nu}_{4q}(x_1,\ldots,x_4),
\label{eq:nonlocal-mol-cur_Zbprime}\\
J^{\mu\nu}_{Z_b'; 4q}&=&  \varepsilon^{\mu\nu\alpha\beta}  \, 
 (\bar d(x_3) \gamma_\alpha   b(x_1))  (\bar b(x_2) \gamma_\beta u(x_4))
\nonumber
\ena 

The effective interaction Lagrangians describing the coupling
of the $Z_b$ and $Z_b^\prime$ states 
to its constituent quarks is written in the form
\bea
{\cal L}_{ {\rm int}, Z_b} &=& 
g_{Z_b}\,Z_{b,\,\mu}(x)\cdot J^\mu_{Z_b}(x) + \text{H.c.}
\label{eq:Zb-lag}
\\     
{\cal L}_{ {\rm int}, Z'_b} &=& \frac{g_{Z_b^\prime}}{2 M_{Z_b^\prime}} \, 
Z_{b,\,\mu\nu}^\prime(x) \cdot J^{\mu\nu}_{Z_b^\prime}(x) + \text{H.c.}
\label{eq:Zb1-lag}\,, 
\ena     
where $Z_{b,\,\mu\nu}^\prime = \partial_\mu  Z_{b,\,\nu}^\prime - 
                               \partial_\nu  Z_{b,\,\mu}^\prime$ is 
the stress tensor of the $Z_b^\prime$ field. We have included a factor $1/M_{Z_b^\prime}$ 
in the interaction Lagrangian for the $Z_{b,\,\mu}$ state to have the same 
dimension for the $g_{Z_b}$ and $g_{Z_b^\prime}$ couplings. 

The coupling constants $g_{H}$, $H = Z_b, Z_b^\prime$ 
in Eqs.~(\ref{eq:Zb-lag}) and (\ref{eq:Zb1-lag}) 
are determined by the normalization condition called 
{\it the compositeness condition}
(see, Refs.~\cite{Efimov:1993ei} and~\cite{Efimov:1988yd} for details).
\be
\label{eq:Z=0}
Z_{H} = 1-g^2_{H}\,\widetilde\Pi_{H}^\prime(M^2_{H})=0,
\en
where $\Pi_{H}(p^2)$ is the scalar part of the vector-meson mass 
operator 
\bea
\widetilde\Pi^{\mu\nu}_{H}(p) &=& g^{\mu\nu} \widetilde\Pi_{H}(p^2)
                            + p^\mu p^\nu \widetilde\Pi^{(1)}_{H}(p^2),\nn
\widetilde\Pi_{H}(p^2) &=&
\frac13\left(g_{\mu\nu}-\frac{p_\mu p_\nu}{p^2}\right)\Pi^{\mu\nu}_{H}(p).
\label{eq:mass}
\ena

The Fourier-transforms of the $Z_b$ and $Z_B^\prime$ mass operators  
are given by
\bea
\widetilde\Pi_{Z_b}^{\mu\nu}(p) &=& 
\frac{9}{2}\,\prod\limits_{i=1}^3\int\!\!\frac{d^4k_i}{(2\pi)^4i}\,
\widetilde\Phi_{Z_b}^2\left(-\,\vec\omega^{\,2}\right) 
\nn
&\times& \Big\{
\,\,\Tr\left[\gamma_5  S_1(\hat k_1)\gamma_5 S_3(\hat k_3)  \right]
    \Tr\left[\gamma^\mu S_4(\hat k_4)\gamma^\nu S_2(\hat k_2)\right] 
\nn
&&
+\, \Tr\left[\gamma^\mu S_1(\hat k_1)\gamma^\nu S_3(\hat k_3)  \right]
    \Tr\left[\gamma_5 S_4(\hat k_4)\gamma_5 S_2(\hat k_2)\right] 
\Big\}
\label{eq:Zb-mass}
\ena
and 
\bea
\widetilde\Pi_{Z_b^\prime}^{\mu\nu}(p) &=& 
-\,9\,\prod\limits_{i=1}^3
\int\!\!\frac{d^4k_i}{(2\pi)^4i}\,
\widetilde\Phi_{Z_b^\prime}^2\left(-\,\vec\omega^{\,2}\right) \ 
\frac{\varepsilon^{\mu p \alpha\beta}  \,  
\varepsilon^{\nu p \rho\sigma}}{M_{Z_b^\prime}^2}
\nn 
&\times& 
\,\,\Tr\left[\gamma_\rho S_1(\hat k_1)\gamma_\alpha S_3(\hat k_3)  \right]
    \Tr\left[\gamma_\beta S_4(\hat k_4)\gamma_\sigma S_2(\hat k_2)\right] 
\label{eq:Zb1-mass} \,, 
\ena 
where 
\bea 
& &\hat k_1=k_1-w_1 p\,, \ \hat k_2=k_2-w_2 p\,, \ \hat k_3=k_3+w_3 p\,, \
k_4=k_1+k_2-k_3+w_4 p\,, \nn 
& &\vec\omega^{\,2}=1/2\,(k_1^2+k_2^2+k_3^2+k_1k_2-k_1k_3-k_2k_3) 
\ena 
and $\varepsilon^{\mu p \alpha\beta} = p_\nu \, 
\varepsilon^{\mu\nu \alpha\beta}$. 
  
The matrix elements of the two-body decays are given by 
\bea
&&
%
%----------------- Zb->Upsi+pi ----------------------------
%
M^{\mu\delta}\left( Z_b(p,\mu) \to \Upsilon(q_1,\delta)+\pi^+(q_2)\right)
= \frac{3}{\sqrt{2}}\,g_{Z_b}g_{\Upsilon}g_{\pi}
\nn
&\times&
\int\!\!\frac{d^4k_1}{(2\pi)^4i}\,\int\!\!\frac{d^4k_2}{(2\pi)^4i}\,
\widetilde\Phi_{Z_b}\left(-\,\vec\eta^{\,2}\right)
\widetilde\Phi_{\Upsilon}\left(-\,(k_1+v_1 q_1)^2\right)
\widetilde\Phi_{\pi}\left(-\,(k_2+u_4 q_2)^2\right)
\nn
&\times& \Big\{
\,\,\Tr\left[ \gamma_5   S_1(k_1)\gamma^\delta S_2(k_1+q_1) 
              \gamma^\mu S_4(k_2)\gamma_5 S_3(k_2+q_2) \right] 
\nn
&&
+\,\Tr\left[ \gamma^\mu   S_1(k_1)\gamma^\delta S_2(k_1+q_1) 
              \gamma_5 S_4(k_2)\gamma_5 S_3(k_2+q_2) \right] 
\Big\}
\nn[2ex]
&=& A_{Z_b\Upsilon\pi}\,g^{\mu\delta} + B_{Z_b\Upsilon\pi}\,q_1^\mu q_2^\delta\,,
\label{eq:decay-mol-b1}
\ena
\bea
&&
%
%----------------- Zb1->Upsi+pi ----------------------------
%
M^{\mu\delta}
\left( Z_b'(p,\mu) \to \Upsilon(q_1,\delta)+\pi^+(q_2)\right)
= 3\,g_{Z_b'}g_{\Upsilon}g_{\pi} \, 
\frac{i\varepsilon^{\mu p \alpha\beta}}{M_{Z_b^\prime}}
\nn
&\times&
\int\!\!\frac{d^4k_1}{(2\pi)^4i}\,\int\!\!\frac{d^4k_2}{(2\pi)^4i}\,
\widetilde\Phi_{Z_b'}\left(-\,\vec\eta^{\,2}\right)
\widetilde\Phi_{\Upsilon}\left(-\,(k_1+v_1 q_1)^2\right)
\widetilde\Phi_{\pi}\left(-\,(k_2+u_4 q_2)^2\right)
\nn
&\times& 
\,\,\Tr\left[ \gamma_\alpha S_1(k_1)\gamma^\delta S_2(k_1+q_1) 
              \gamma_\beta S_4(k_2)\gamma_5 S_3(k_2+q_2) \right] 
\nn[2ex]
&=& A_{Z_b'\Upsilon\pi}\,g^{\mu\delta} + B_{Z_b'\Upsilon\pi}\,q_1^\mu q_2^\delta\,,
\label{eq:decay-mol-b1prime}
\ena
\bea 
&&
%
%----------------- Zb->etab+rho ----------------------------
%
M^{\mu\rho}
\left( Z_b(p,\mu) \to \eta_b(q_1)+\rho(q_2,\rho)\right)
= \frac{3}{\sqrt{2}}\,g_{Z_b}g_{\eta_b}g_{\rho}
\nn
&\times&
\int\!\!\frac{d^4k_1}{(2\pi)^4i}\,\int\!\!\frac{d^4k_2}{(2\pi)^4i}\,
\widetilde\Phi_{Z_b}\left(-\,\vec\eta^{\,2}\right)
\widetilde\Phi_{\eta_b}\left(-\,(k_1+v_1 q_1)^2\right)
\widetilde\Phi_{\rho}\left(-\,(k_2+u_4 q_2)^2\right)
\nn
&\times& \Big\{
\,\,\Tr\left[ \gamma_5 S_1(k_1)\gamma_5 S_2(k_1+q_1)
              \gamma^\mu S_4(k_2)\gamma^\rho S_3(k_2+q_2) \right]
\nn
&&
+\,\Tr\left[  \gamma^\mu S_1(k_1)\gamma_5 S_2(k_1+q_1)
              \gamma_5 S_4(k_2)\gamma^\rho S_3(k_2+q_2) \right]
\Big\}
\nn[2ex]
&=& A_{Z_b\eta_b\rho}\,g^{\mu\rho} - B_{Z_b\eta_b\rho}\,q_2^\mu q_1^\rho\,.
\label{eq:decay-mol-Zbetabrho}
\ena
\bea
&&
%
%----------------- Zb1->etab+pi ----------------------------
%
M^{\mu\rho}\left( Z_b'(p,\mu) \to \eta_b(q_1) +\rho(q_2,\rho)\right)
= 3 \,g_{Z_b'}g_{\eta_b}g_{\rho} \,
\frac{i\varepsilon^{\mu p \alpha\beta}}{M_{Z_b^\prime}}
\nn
&\times&
\int\!\!\frac{d^4k_1}{(2\pi)^4i}\,\int\!\!\frac{d^4k_2}{(2\pi)^4i}\,
\widetilde\Phi_{Z_b'}\left(-\,\vec\eta^{\,2}\right)
\widetilde\Phi_{\eta_b}\left(-\,(k_1+v_1 q_1)^2\right)
\widetilde\Phi_{\rho}\left(-\,(k_2+u_4 q_2)^2\right)
\nn
&\times&
\,\,\Tr\left[ \gamma_{\alpha} S_1(k_1)\gamma_5 S_2(k_1+q_1)
              \gamma_\beta  S_4(k_2)\gamma^\rho S_3(k_2+q_2) \right]
\nn[2ex]
&=& A_{Z_b'\eta_b\rho}\,g^{\mu\rho} - B_{Z_b'\eta_b\rho}\,q_2^\mu q_1^\rho\,.
\label{eq:decay-mol-Zbprimeetavrho}
\ena
\bea
&&
%
%----------------- Zb->hb+pi ----------------------------
M^{\mu\delta}
\left( Z_b^+(p,\mu) \to h_b(q_1,\delta)
+\pi^+(q_2)\right)
= \frac{3}{\sqrt{2}}\,g_{Z_b}g_{h_b}g_{\pi} \, 
\nn 
&\times&\int\!\!\frac{d^4k_1}{(2\pi)^4i}\,\int\!\!\frac{d^4k_2}{(2\pi)^4i}\,
\widetilde\Phi_{Z_b}\left(-\,\vec\eta^{\,2}\right)
\widetilde\Phi_{h_b}\left(-\,(k_1+v_1 q_1)^2\right)
\widetilde\Phi_{\pi}\left(-\,(k_2+u_4 q_2)^2\right)
\nn 
&\times&  \Big\{
\,\,\Tr\left[ \gamma_5   S_1(k_1) \gamma_5\cdot (2 k_1^\delta) S_2(k_1+q_1) 
              \gamma^\mu S_4(k_2)\gamma_5 S_3(k_2+q_2) \right] 
\nn
&&
+\,\Tr\left[ \gamma^\mu   S_1(k_1) \gamma_5\cdot (2 k_1^\delta) S_2(k_1+q_1) 
              \gamma_5 S_4(k_2)\gamma_5 S_3(k_2+q_2) \right] 
\Big\}
\nn[2ex]
&=& \varepsilon^{\mu\delta q_1 q_2} A_{Z_bh_b\pi}\,, 
\label{eq:decay-mol-b2}
\ena
\bea
&&
%
%----------------- Zb1->hb+pi ----------------------------
%
M^{\mu\delta}
\left( Z_b'(p,\mu) \to h_b(q_1,\delta)+\pi^+(q_2)\right)
= 3 \,g_{Z_b'}g_{h_b}g_{\pi} \, 
\frac{i\varepsilon^{\mu p \alpha\beta}}{M_{Z_b^\prime}}
\nn
&\times&
\int\!\!\frac{d^4k_1}{(2\pi)^4i}\,\int\!\!\frac{d^4k_2}{(2\pi)^4i}\,
\widetilde\Phi_{Z_b'}\left(-\,\vec\eta^{\,2}\right)
\widetilde\Phi_{h_b}\left(-\,(k_1+v_1 q_1)^2\right)
\widetilde\Phi_{\pi}\left(-\,(k_2+u_4 q_2)^2\right)
\nn
&\times& 
\,\,\Tr\left[ \gamma_\alpha S_1(k_1)\gamma_5\cdot(2 k_1^\delta) S_2(k_1+q_1) 
              \gamma_\beta S_4(k_2)\gamma_5 S_3(k_2+q_2) \right] 
\nn[2ex]
&=& \varepsilon^{\mu\delta q_1 q_2} A_{Z_b'h_b\pi}\,. 
\label{eq:decay-mol-b2prime}
\ena
The argument $\vec\eta^{\,2}$ of the 
$Z_b$ and $Z_b^\prime$ vertex functions is given by 
\bea
\vec\eta^{\,2} &=& \eta_1^2+\eta_2^2+\eta_3^2,
\nn
\eta_1 &=& + \frac{1}{2\sqrt{2}}\left(2k_1+(1+w_1-w_2) q_1 +(w_1-w_2) q_2 
\right),
\nn
\eta_2 &=& + \frac{1}{2\sqrt{2}}\left(2k_2-(w_3-w_4) q_1 +(1-w_3+w_4) q_2 
\right),
\nn
\eta_3 &=& + \frac{1}{2}\left((1-w_1-w_2) q_1 - (w_1+w_2) q_2 \right)\,,
\label{eq:Zc-arg-CC}
\ena 
where $w_i = m_i/(m_1+m_2+m_3+m_4)$. 
The quark masses $m_i$ are specified as 
$m_1=m_2=m_b$,  $m_3=m_4=m_d=m_u$,
and the two-body reduced masses as $v_i=m_i/(m_1+m_2)$ $(i=1,2)$ 
and $u_j=m_j/(m_3+m_4)$ $(j=3,4)$.

The matrix elements of the decays 
$Z^+_b\to \bar B^{0}+B^{\ast\,+}$ and 
$Z^+_b\to \bar B^{\ast\, 0}+B^{+}$ read 
\bea
&&
M^{\mu\rho}
\left( Z_b^+(p,\mu) \to\bar B^0(q_1)+B^{\ast\,+}(q_2,\rho)\right)
=\frac{9}{\sqrt{2}}\,g_{Z_b}g_{B}g_{B^\ast}
\nn
&\times&
\int\!\!\frac{d^4k_1}{(2\pi)^4i}\,\int\!\!\frac{d^4k_2}{(2\pi)^4i}\,
\widetilde\Phi_{Z_b}\left(-\,\vec\delta^{\,2}\right)
   \widetilde\Phi_{B}\left(-\,(k_2+v_4 q_1)^2\right)
\widetilde\Phi_{B^\ast}\left(-\,(k_1+u_1 q_2)^2\right)
\nn
&\times& 
\,\,\Tr\left[ \gamma^\mu S_1(k_1)\gamma^\rho S_3(k_1+q_2)  \right]  
    \Tr\left[ \gamma_5 S_4(k_2)\gamma_5 S_2(k_2+q_1)  \right] 
\nn[2ex]
&=& A_{Z_b\bar BB^\ast}\,g^{\mu\rho} - B_{Z_b\bar BB^\ast}\,q_2^\mu q_1^\rho\,,
\label{eq:decay-mol-BB1}
\ena

\bea
&&
M^{\mu\alpha}\left( Z_b^+(p,\mu) \to\bar B^{\ast\,0} (q_1,\delta)+B^+(q_2)\right)
=\frac{9}{\sqrt{2}}\,g_{Z_b}g_{B^\ast}g_{B}
\nn
&\times&
\int\!\!\frac{d^4k_1}{(2\pi)^4i}\,\int\!\!\frac{d^4k_2}{(2\pi)^4i}\,
\widetilde\Phi_{Z_b}\left(-\,\vec\delta^{\,2}\right)
\widetilde\Phi_{B^\ast}\left(-\,(k_1+\hat v_1 q_1)^2\right)
\widetilde\Phi_{B}\left(-\,(k_2+\hat u_4 q_2)^2\right)
\nn
&\times& 
\,\,\Tr\left[\gamma_5   S_1(k_1) \gamma_5    S_3(k_1+q_2) \right] 
    \Tr\left[\gamma^\mu S_4(k_2) \gamma^\delta S_2(k_2+q_1) \right] 
\nn[2ex]
&=& A_{Z_bB^\ast B}\,g^{\mu\delta} + B_{Z_bB^\ast B}\,q_1^\mu q_2^\delta\,.
\label{eq:decay-mol-DD2}
\ena

The argument of $Z_b$-vertex function is given by
\bea
\vec\delta^{\,2} &=& \delta_1^2+\delta_2^2+\delta_3^2,
\nn
\delta_1 &=& - \frac{1}{2\sqrt{2}}
\left(k_1 + k_2 + (w_1-w_2) q_1 + (1+w_1-w_2)q_2) \right),
\nn
\delta_2 &=& + \frac{1}{2\sqrt{2}}
\left(k_1 + k_2+(1-w_3+w_4) q_1 - (w_3-w_4) q_2 \right),
\nn
\delta_3 &=& + \frac{1}{2}
\left(k_1 - k_2 + (w_1+w_2) q_1 - (1-w_1-w_2)q_2) \right).
\label{eq:arg-mol-DD2}
\ena
The quark masses are specified as $m_1=m_2=m_b$,  $m_3=m_4=m_d=m_u$,
and the two-body reduced masses as
$\hat v_2=m_2/(m_2+m_4)$,  $\hat v_4=m_4/(m_2+m_4)$
and $\hat u_1=m_1/(m_1+m_3)$, $\hat u_3=m_3/(m_1+m_3)$. 

Finally, we consider the 
$Z_b^{\prime +} \to B^{\ast\,+} + \bar B^{\ast\,0}$ decay. 
This process is described by the invariant matrix element, 
which is expressed in terms of three relativistic amplitudes 
$B_i$, $(i=1,2,3)$ as  
\bea 
&&
M^{\mu\delta\rho} (Z_b^{\prime +}(p,\mu) \to 
B^{\ast 0}(q_1,\delta) + \bar B^{\ast +}(q_2,\rho)) 
= 9\,g_{Z_b'}g_{B^\ast}g_{B^\ast} \, 
\frac{\varepsilon^{\mu p \alpha\delta} }{M_{Z_b^\prime}}
\nn
&\times&
\int\!\!\frac{d^4k_1}{(2\pi)^4i}\,\int\!\!\frac{d^4k_2}{(2\pi)^4i}\,
\widetilde\Phi_{Z_b'}\left(-\,\vec\delta^{\,2}\right)
\widetilde\Phi_{B^\ast}\left(-\,(k_1+\hat v_1 q_1)^2\right)
\widetilde\Phi_{B^\ast}\left(-\,(k_2+\hat u_4 q_2)^2\right)
\nn
&\times& 
\,\,\Tr\left[\gamma_\alpha S_1(k_1) \gamma^\delta S_3(k_1+q_1) \right] 
    \Tr\left[\gamma_\beta S_4(k_2) \gamma^\rho S_2(k_2+q_2) \right] 
\nn[2ex]
&=& 
  B_1 q^\mu_1 \epsilon^{q_1q_2\rho\delta} 
+ B_2 \epsilon^{q_1\mu\rho\delta}
+ B_3 \epsilon^{q_2\mu\rho\delta} \,.
\ena 

\section{Numerical results}
\label{sec:numerics}

First of all, we would like to note that all adjustable parameters
of our model (constituent quark masses, infrared cut-off and size
parameters) have been fixed in our previous studies by a global fit to
a multitude of experimental data~\cite{Olive:2016xmw}. 
The only two new parameters are the size parameters of the two exotic
$Z_b(Z'_b)$ states. As a guide to adjust them  we take the experimental
values of
the largest branching fractions presented in Ref.~\cite{Garmash:2015rfd}:
\bea
{\cal B}(Z_b^+\to [B^+ \bar B^{\ast\,0} + \bar B^0 B^{\ast\,+}]) &=& 85.6^{+1.5+1.5}_{-2.0-2.1}\,\%\,,
\nn[2ex]    
{\cal B}(Z_b^{\prime +} \to \bar B^{\ast\,+} B^{\ast\,0}) &=& 73.7^{+3.4+2.7}_{-4.4-3.5}\,\%\,.
\label{eq:expt-Br}
\ena
By using the central values of these branching rates and total decay widths
given in Eq.~(\ref{eq:Belle-2}) we find the central values of our
size parameters $\Lambda_{Z_b}=3.45$~GeV and  $\Lambda_{Z'_b}=3.00$~GeV.
Allowing them to vary in the interval
\be
\Lambda_{Z_b} =3.45\pm 0.05\,\,\text{GeV} \qquad
\Lambda_{Z'_b}=3.00\pm 0.05\,\,\text{GeV}\,,
\label{eq:Lambda}
\en
we obtain the values of various decay widths shown in Table~\ref{tab:theory}.
\begin{table}[!t]
\centering
\caption{Particle decay widths for the $Z^+_b(10610)$ and $Z^+_b(10650)$.} 
\medskip
\label{tab:theory}
\def\arraystretch{1.5}
  \begin{tabular}{lll}  
\hline \hline
  ~Channel~\hspace*{5mm}  & \multicolumn{2}{c}{Widths, MeV}   \\
              & \qquad $Z_b(10610)$ \qquad  & \qquad  $Z_b(10650)$ \qquad      
\\
\hline 
 $\Upsilon(1S)\pi^+$ & \qquad $5.9\pm 0.4$   & \qquad $9.5^{+0.7}_{-0.6}$ 
\\
 $h_b(1P)\pi^+$ & \qquad $(0.14\pm 0.01)\cdot 10^{-1}$ & \qquad
$0.74^{+0.05}_{-0.04}\cdot 10^{-3}$
\\
 $\eta_b \rho^+$  & \qquad $4.4\pm 0.3$ & \qquad $7.5^{+0.6}_{-0.5}$ 
\\
$B^+\bar{B}^{*0}+\bar{B}^0B^{*+}$ &  \qquad $20.7^{+1.6}_{-1.5}$    &
\qquad $-$         
\\
 $B^{*+}\bar{B}^{*0}$              &  \qquad $-$  & \qquad $17.1^{+1.5}_{-1.4}$  
\\
\hline \hline
  \end{tabular}
\end{table}

The total widths are equal to $\sum_i\Gamma_i(Z_b)=30.9^{+2.3}_{-2.1}.0$~MeV
and  $\sum_i\Gamma_i(Z'_b)=34.1^{+2.8}_{-2.5}$~MeV which should be compared with
the experimental values  $\Gamma(Z_b)=25 \pm 7$~MeV
and $\Gamma(Z'_b)=23 \pm 8$~MeV, respectively.
The Belle observations indicate that the decays involving  
bottomonium states are significantly suppressed
compared with the $B$-meson modes. In our calculation we find that 
the modes with $\Upsilon(1S)\pi^+$  and $\eta_b \rho^+$ 
are suppressed but not as much as in the data.  As one can see from 
Table~\ref{tab:theory} the ratios of decay rates are
\bea
\frac{\Gamma\left( Z_b\to\Upsilon(1S)\pi\right)}
     {\Gamma\left( Z_b\to B \bar{B}^\ast + c.c. \right)}
&\approx& 0.29\,, \qquad
     \frac{\Gamma\left( Z_b\to\eta_b\rho \right)}
          {\Gamma\left( Z_b\to B \bar B^\ast + c.c. \right)}
\approx 0.21\,,
\nn
\frac{\Gamma\left( Z'_b\to\Upsilon(1S)\pi \right)}
     {\Gamma\left( Z'_b\to B^\ast \bar B^\ast \right)}
&\approx& 0.56\,, \qquad\qquad
     \frac{\Gamma\left( Z'_b\to\eta_b\rho \right)}
          {\Gamma\left( Z'_b\to B^\ast \bar B^\ast \right)}
\approx 0.44\,.
\label{eq:ratios}
\ena

The decays into the $h_b(1P)\pi^+$ mode 
are suppressed by the $p$-wave suppression factor in the
rate expression~(\ref{eq:width1p0m}).

\section{Summary}
\label{sec:Summary}

By using molecular-type four-quark currents for the recently
observed resonances $Z_b(10610)$ and $Z_b(10650)$, we have calculated
their two-body decay rates into a bottomonium state plus a light meson
as well as  into $B$-meson pairs.

We have fixed the model size parameters by adjusting
the theoretical values of the largest branching fractions
of the modes with the $B$-mesons in the final states to their
experimental values.

We found that the modes with $\Upsilon(1S)\pi^+$  and $\eta_b \rho^+$ 
in the final states are suppressed but not as much as the Belle Collaboration
reported.

\begin{acknowledgments}

This work was supported
by the German Bundesministerium f\"ur Bildung und Forschung (BMBF)
under Project 05P2015 - ALICE at High Rate (BMBF-FSP 202):
``Jet- and fragmentation processes at ALICE and the parton structure 
of nuclei and structure of heavy hadrons'', 
by CONICYT (Chile) PIA/Basal FB0821, 
by Tomsk State University Competitiveness 
Improvement Program and the Russian Federation program ``Nauka'' 
(Contract No. 0.1764.GZB.2017). 
The research is carried out at Tomsk Polytechnic University within the framework 
of Tomsk Polytechnic University Competitiveness Enhancement Program grant.
M.A.I.\ acknowledges the support from  PRISMA cluster of excellence 
(Mainz Uni.). M.A.I. and J.G.K. thank the Heisenberg-Landau Grant for
partial support.  

\end{acknowledgments}


\begin{thebibliography}{99}

%\cite{Belle:2011aa}
\bibitem{Belle:2011aa} 
  A.~Bondar {\it et al.} (Belle Collaboration),
  %``Observation of two charged bottomonium-like resonances in Y(5S) decays,''
  Phys.\ Rev.\ Lett.\  {\bf 108}, 122001 (2012).
%  doi:10.1103/PhysRevLett.108.122001
%  [arXiv:1110.2251 [hep-ex]].

%\cite{Garmash:2014dhx}
\bibitem{Garmash:2014dhx} 
  A.~Garmash {\it et al.} (Belle Collaboration),
  %``Amplitude analysis of $e^+e^- \to \Upsilon(nS) \pi^+\pi^-$ at 
  %$\sqrt{s}=10.865$~GeV,''
  Phys.\ Rev.\ D {\bf 91}, 072003 (2015).
%  doi:10.1103/PhysRevD.91.072003
%  [arXiv:1403.0992 [hep-ex]].

%\cite{Garmash:2015rfd}
\bibitem{Garmash:2015rfd} 
  A.~Garmash {\it et al.} (Belle Collaboration),
  %``Observation of Zb(10610) and Zb(10650) Decaying to B Mesons,''
  Phys.\ Rev.\ Lett.\  {\bf 116}, no. 21, 212001 (2016). 
%  doi:10.1103/PhysRevLett.116.212001
%  [arXiv:1512.07419 [hep-ex]].

\bibitem{Bondar:2011ev} 
  A.~E.~Bondar, A.~Garmash, A.~I.~Milstein, R.~Mizuk and M.~B.~Voloshin,
  %``Heavy quark spin structure in $Z_b$ resonances,''
  Phys.\ Rev.\ D {\bf 84}, 054010 (2011).
%  doi:10.1103/PhysRevD.84.054010
%  [arXiv:1105.4473 [hep-ph]].

\bibitem{Voloshin:2011qa}
  M.~B.~Voloshin,
  %``Radiative transitions from Upsilon(5S) to molecular bottomonium,''       
  Phys.\ Rev.\  D {\bf 84}, 031502 (2011).
%  [arXiv:1105.5829 [hep-ph]].

\bibitem{Mqh:2011}
  J.~R.~Zhang, M.~Zhong and M.~Q.~Huang,
  %``Could $Z_{b}(10610)$ be a $B^{*}\bar{B}$ molecular state?,''             
  Phys.\ Lett.\  B {\bf 704}, 312 (2011).
%  [arXiv:1105.5472 [hep-ph]];
  %%CITATION = PHLTA,B704,312;%%                                              
  C.~Y.~Cui, Y.~L.~Liu, M.~Q.~Huang,
  %``Investigating different structures for the Z_{b}(10610)                   
  %and Z_{b}(10650),                                                        
  Phys.\ Rev.\  D {\bf 85}, 074014 (2012). 
%  [arXiv:1107.1343 [hep-ph]].
  %%CITATION = PHRVA,D85,074014;%% 
%                                                                             
 \bibitem{Pjl:2011} 
 Y.~Yang, J.~Ping, C.~Deng and H.~S.~Zong,
  %``Dynamical study of the $Z_b$(10610) and $Z_b$(10650)                      
  %as molecular states,''                                                     
  J.\ Phys.\ G {\bf 39}, 105001 (2012). 
%  [arXiv:1105.5935 [hep-ph]].
  %%CITATION = ARXIV:1105.5935;%%                                           
%                                                                             
 %\cite{Danilkin:2011sh}                                                      
\bibitem{Danilkin:2011sh}
  I.~V.~Danilkin, V.~D.~Orlovsky and Yu.~A.~Simonov,
  %``Hadron interaction with heavy quarkonia,''     
  Phys.\ Rev.\  D {\bf 85}, 034012 (2012).
%  [arXiv:1106.1552 [hep-ph]].
  %%CITATION = PHRVA,D85,034012;%%                                            
%                                                                             
 \bibitem{Slzhu:2011}
  Z.~F.~Sun, J.~He, X.~Liu, Z.~G.~Luo and S.~L.~Zhu,
  %``$Z_b(10610)^\pm$ and $Z_b(10650)^\pm$ as the $B^*\bar{B}$ and          
  %$B^*\bar{B}^{*}$ molecular states,''                                      
  Phys.\ Rev.\  D {\bf 84}, 054002 (2011). 
%  [arXiv:1106.2968 [hep-ph]].

\bibitem{Dyc:2011}
  D.~Y.~Chen, X.~Liu and S.~L.~Zhu,
  %``Charged bottomonium-like states $Z_b(10610)$ and $Z_b(10650)$ and the 
  %$\Upsilon(5S)\to \Upsilon(2S)\pi^+\pi^-$ decay,''                          
  Phys.\ Rev.\  D {\bf 84}, 074016 (2011);
%  [arXiv:1105.5193 [hep-ph]];
  %%CITATION = PHRVA,D84,074016;%%                                            
  D.~Y.~Chen and X.~Liu,
  %``$Z_b(10610)$ and $Z_b(10650)$ structures produced by the initial single  
  %pion emission in the $\Upsilon(5S)$ decays,''                          
  Phys.\ Rev.\  D {\bf 84}, 094003 (2011);
%  [arXiv:1106.3798 [hep-ph]];
  %%CITATION = PHRVA,D84,094003;%%                                            
  X.~Liu and D.~Y.~Chen,
  %``Charged bottomonium-like structures $Z_b(10610)$ and $Z_b(10650)$,''     
  Few Body Syst.\  {\bf 54}, 165 (2013). 
%  [arXiv:1110.4433 [hep-ph]].
  %%CITATION = ARXIV:1110.4433;%%                                        
%                                                                             
  \bibitem{Cleven:2011gp}
  M.~Cleven, F.~K.~Guo, C.~Hanhart and U.~G.~Meissner,
  %``Bound state nature of the exotic Z_b states,''                           
  Eur.\ Phys.\ J.\  A {\bf 47}, 120 (2011). 
%  [arXiv:1107.0254 [hep-ph]].
  %%CITATION = EPHJA,A47,120;%%                                              
%                                                                             
  \bibitem{Guo:2011gu}
  T.~Guo, L.~Cao, M.~Z.~Zhou and H.~Chen,
  %``The possible candidates of tetraquark : $Z_b(10610)$ and $Z_b(10650)$,''
  arXiv:1106.2284 [hep-ph].
  %%CITATION = ARXIV:1106.2284;%%                                             
%                                                                             
  \bibitem{Richard:2011}
  F.~S.~Navarra, M.~Nielsen, J.~-M.~Richard,
  %``Exotic Charmonium and Bottomonium-like Resonances,''                 
  J.\ Phys.\ Conf.\ Ser.\  {\bf 348}, 012007 (2012). 
%  [arXiv:1108.1230 [hep-ph]].
  %%CITATION = 00462,348,012007;%%

%\cite{Li:2012wf}
\bibitem{Li:2012wf} 
  M.~T.~Li, W.~L.~Wang, Y.~B.~Dong and Z.~Y.~Zhang,
  %``$Z_b(10650)$ and $Z_b(10610)$ States in A Chiral Quark Model,''
  J.\ Phys.\ G {\bf 40}, 015003 (2013).
  %doi:10.1088/0954-3899/40/1/015003
%  [arXiv:1204.3959 [hep-ph]].

%\cite{Chen:2012yr}
\bibitem{Chen:2012yr} 
  D.~Y.~Chen, X.~Liu and T.~Matsuki,
  %``Interpretation of $Z_b$(10610) and $Z_b$(10650) in the ISPE mechanism 
  %and the Charmonium Counterpart,''
  Chin.\ Phys.\ C {\bf 38}, 053102 (2014).
  %doi:10.1088/1674-1137/38/5/053102
%  [arXiv:1208.2411 [hep-ph]].

%\cite{Dong:2012hc}
\bibitem{Dong:2012hc} 
  Y.~Dong, A.~Faessler, T.~Gutsche and V.~E.~Lyubovitskij,
  %``Decays of Zb(+) and Zb'(+) as Hadronic Molecules,''
  J.\ Phys.\ G {\bf 40}, 015002 (2013);
  %doi:10.1088/0954-3899/40/1/015002
%  [arXiv:1203.1894 [hep-ph]];
%\cite{Dong:2013rsa}
%\bibitem{Dong:2013rsa} 
%  Y.~Dong, A.~Faessler, T.~Gutsche and V.~E.~Lyubovitskij,
  %``A Study of New Resonances in a Molecule Scenario,''
  Few Body Syst.\  {\bf 54}, 1011 (2013).
%  doi:10.1007/s00601-013-0622-4


\bibitem{Li:2012as} 
  G.~Li, F.~l.~Shao, C.~W.~Zhao and Q.~Zhao,
  %``$Z_b/Z_b^\prime \to \Upsilon\pi$ and $h_b \pi$ decays 
  %in intermediate meson loops model,''
  Phys.\ Rev.\ D {\bf 87}, 034020 (2013).
  %doi:10.1103/PhysRevD.87.034020
%  [arXiv:1212.3784 [hep-ph]].

%\cite{Cleven:2013sq}
\bibitem{Cleven:2013sq} 
  M.~Cleven, Q.~Wang, F.~K.~Guo, C.~Hanhart, U.~G.~Meissner and Q.~Zhao,
  %``Confirming the molecular nature of the $Z_b(10610)$ 
  %and the $Z_b(10650)$,''
  Phys.\ Rev.\ D {\bf 87}, 074006 (2013). 
  %doi:10.1103/PhysRevD.87.074006
%  [arXiv:1301.6461 [hep-ph]].

%\cite{Ohkoda:2013cea}
\bibitem{Ohkoda:2013cea} 
  S.~Ohkoda, S.~Yasui and A.~Hosaka,
  %``Decays of $Z_b \to \Upsilon \pi$ via triangle diagrams 
  %in heavy meson molecules,''
  Phys.\ Rev.\ D {\bf 89}, 074029 (2014).
  %doi:10.1103/PhysRevD.89.074029
%  [arXiv:1310.3029 [hep-ph]].
  
%\cite{Wang:2013zra}
\bibitem{Wang:2013zra} 
  Z.~G.~Wang and T.~Huang,
  %``The $Z_b(10610)$ and $Z_b(10650)$ 
  %as axial-vector tetraquark states in the QCD sum rules,''
  Nucl.\ Phys.\ A {\bf 930}, 63 (2014).
  %doi:10.1016/j.nuclphysa.2014.08.084
%  [arXiv:1312.2652 [hep-ph]].
%%\cite{Wang:2014gwa}
%\bibitem{Wang:2014gwa} 
  Z.~G.~Wang,
  %``Reanalysis of the $Y(3940)$, $Y(4140)$, $Z_c(4020)$, $Z_c(4025)$ 
  %and $Z_b(10650)$ as molecular states with QCD sum rules,''
  Eur.\ Phys.\ J.\ C {\bf 74}, 2963 (2014).
  %doi:10.1140/epjc/s10052-014-2963-7
%  [arXiv:1403.0810 [hep-ph]].

%\cite{Dias:2014pva}
\bibitem{Dias:2014pva} 
  J.~M.~Dias, F.~Aceti and E.~Oset,
  %``Study of $B\bar{B}^*$ and $B^*\bar{B}^*$ interactions in $I=1$ 
  %and relationship to the $Z_b(10610)$, $Z_b(10650)$ states,''
  Phys.\ Rev.\ D {\bf 91}, 076001 (2015).
  %doi:10.1103/PhysRevD.91.076001
%  [arXiv:1410.1785 [hep-ph]].

%\cite{Huo:2015uka}
\bibitem{Huo:2015uka} 
  W.~S.~Huo and G.~Y.~Chen,
  %``The nature of $Z_b$ states from a combined analysis 
  %of $\Upsilon (5S)\rightarrow h_b(mP) \pi ^+ \pi ^-$ and 
  %$\Upsilon (5S)\rightarrow B^{(*)}\bar{B}^{(*)}\pi $,''
  Eur.\ Phys.\ J.\ C {\bf 76}, 172 (2016).
  %doi:10.1140/epjc/s10052-016-4013-0
%  [arXiv:1501.02189 [hep-ph]].

%\cite{Kang:2016ezb}
\bibitem{Kang:2016ezb} 
  X.~W.~Kang, Z.~H.~Guo and J.~A.~Oller,
  %``General considerations on the nature of $Z_b(10610)$ 
  %and $Z_b(10650)$ from their pole positions,''
  Phys.\ Rev.\ D {\bf 94}, 014012 (2016).
  %doi:10.1103/PhysRevD.94.014012
%  [arXiv:1603.05546 [hep-ph]].

%\cite{Gutsche:2017oro}
\bibitem{Gutsche:2017oro} 
  T.~Gutsche, V.~E.~Lyubovitskij and I.~Schmidt,
  %``Tetraquarks in holographic QCD,''
  arXiv:1706.07716 [hep-ph].

%\cite{hmappr}
\bibitem{hmappr}
  A.~Faessler, T.~Gutsche, V.~E.~Lyubovitskij and Y.~L.~Ma,
  %``Strong and radiative decays of the D(s0)*(2317) meson in the DK-molecule picture,''                                                         
  Phys.\ Rev.\ D {\bf 76}, 014005 (2007); 
%  [arXiv:0705.0254 [hep-ph]];
%  A.~Faessler, T.~Gutsche, V.~E.~Lyubovitskij and Y.~L.~Ma,
  %``D* K molecular structure of the D(s1)(2460) meson,''                                                                                        
  Phys.\ Rev.\ D {\bf 76}, 114008 (2007);
%  [arXiv:0709.3946 [hep-ph]];
%  A.~Faessler, T.~Gutsche, V.~E.~Lyubovitskij and Y.~L.~Ma,
  %``Molecular structure of the B*(sl)(5725) and B(s1)(5778) bottom-strange mesons,''                                                            
  Phys.\ Rev.\ D {\bf 77}, 114013 (2008);
%  [arXiv:0801.2232 [hep-ph]];
  A.~Faessler, T.~Gutsche, S.~Kovalenko and V.~E.~Lyubovitskij,
  %``D/s0*(2317) and D/s1(2460) mesons in two-body B-meson decays,''                                                                             
  Phys.\ Rev.\  D {\bf 76}, 014003 (2007);
%  [arXiv:0705.0892 [hep-ph]];
  Y.~Dong, A.~Faessler, T.~Gutsche, and V.~E.~Lyubovitskij,
  %Estimate for the $X(3872) \to \gamma J/\psi$ decay width,                                                                                      
  Phys.\ Rev.\ D {\bf 77}, 094013 (2008); 
%  Y.~Dong, A.~Faessler, T.~Gutsche and V.~E.~Lyubovitskij,
  %``J/psi gamma and psi(2S) gamma decay modes of the X(3872),''
  J.\ Phys.\ G {\bf 38}, 015001 (2011);
%  [arXiv:0909.0380 [hep-ph]];
%  Y.~Dong, A.~Faessler, T.~Gutsche and V.~E.~Lyubovitskij,
  %``Decays of Zb(+) and Zb'(+) as Hadronic Molecules,''
  J.\ Phys.\ G {\bf 40}, 015002 (2013); 
%  [arXiv:1203.1894 [hep-ph]].
%  Y.~Dong, A.~Faessler, T.~Gutsche and V.~E.~Lyubovitskij,
  %``Strong decays of molecular states Z$_{c}^{+}$ and Z$_{c}^{'+}$,''
  Phys.\ Rev.\ D {\bf 88}, 014030 (2013); 
%  [arXiv:1306.0824 [hep-ph]].
  T.~Branz, T.~Gutsche and V.~E.~Lyubovitskij,
  %``Weak decays of heavy hadron molecules involving the f0(980),''                                                                              
  Phys.\ Rev.\  D {\bf 79}, 014035 (2009);
%  [arXiv:0812.0942 [hep-ph]];
%  T.~Branz, T.~Gutsche and V.~E.~Lyubovitskij,
  %``Hadronic Molecule Structure Of The Y(3940) And Y(4140),''                                                                                   
  Phys.\ Rev.\  D {\bf 80}, 054019 (2009); 
  Y.~Dong, A.~Faessler, T.~Gutsche, Sergey Kovalenko, and V.~E.~Lyubovitskij,
  %X(3872) as a hadronic molecule and its decays to charmonium states and pions,                                                                 
  Phys.\ Rev.\ D {\bf 79}, 094013 (2009); 
%  [arXiv:0903.5424 [hep-ph]];
  T.~Branz, T.~Gutsche and V.~E.~Lyubovitskij,
  %``Two-photon decay of heavy hadron molecules,''                                                                                               
  Phys.\ Rev.\  D {\bf 82}, 054010 (2010);
%  [arXiv:1007.4311 [hep-ph]];
%  T.~Branz, T.~Gutsche and V.~E.~Lyubovitskij,
  %``Hidden-charm and radiative decays of the Z(4430) as a hadronic $D_{1}                                                                       
  %\overline{D^\ast}$ bound state,''                                                                                                             
  Phys.\ Rev.\  D {\bf 82}, 054025 (2010); 
%  [arXiv:1005.3168 [hep-ph]];
  Y.~Dong, A.~Faessler, T.~Gutsche, S.~Kumano and V.~E.~Lyubovitskij,
  %``Radiative decay of Lambdac(2940)+ in a hadronic molecule picture,''                                                                         
  Phys.\ Rev.\  D {\bf 82}, 034035 (2010);
%  [arXiv:1006.4018 [hep-ph]];
  T.~Gutsche, M.~Kesenheimer and V.~E.~Lyubovitskij,
  %Radiative and dilepton decays of the hadronic molecule $Z_c^+$(3900),
  Phys.\ Rev.\ D {\bf 90}, 094013 (2014);
  %[arXiv:1410.0259 [hep-ph]].            
%\bibitem{Dong:2017rmg} 
  Y.~Dong, A.~Faessler, T.~Gutsche, Q.~L\"u and V.~E.~Lyubovitskij,
  %``Selected strong decays of $\eta(2225)$ and $\phi(2170)$ 
  %as $\Lambda \bar\Lambda$ bound states,''
  arXiv:1705.09631 [hep-ph].

%\cite{Weinberg:1962hj}                                                     
\bibitem{Weinberg:1962hj}
  S.~Weinberg,
  %``Elementary Particle Theory Of Composite Particles,''                 
  Phys.\ Rev.\  {\bf 130}, 776 (1963);
  A.~Salam,
  %``Lagrangian Theory Of Composite Particles,''                           
  Nuovo Cim.\  {\bf 25}, 224 (1962);
  K.~Hayashi, M.~Hirayama, T.~Muta, N.~Seto and T.~Shirafuji,
  Fortsch.\ Phys.\ {\bf 15}, 625 (1967).
%\cite{Efimov:1993ei}                                                        
\bibitem{Efimov:1993ei}
  G.~V.~Efimov and M.~A.~Ivanov,
  {\it The Quark Confinement Model of Hadrons},
  (IOP Publishing, Bristol $\&$ Philadelphia, 1993).

%------------------------------------------------

%\cite{Dubnicka:2010kz} 
\bibitem{Dubnicka:2010kz} 
  S.~Dubnicka, A.~Z.~Dubnickova, M.~A.~Ivanov and J.~G.~K\"orner,
  %``Quark model description of the tetraquark state X(3872) 
  % in a relativistic constituent quark model with infrared confinement,''
  Phys.\ Rev.\ D {\bf 81}, 114007 (2010). 
%  doi:10.1103/PhysRevD.81.114007
%  [arXiv:1004.1291 [hep-ph]].

%\cite{Dubnicka:2011mm}
\bibitem{Dubnicka:2011mm} 
  S.~Dubnicka, A.~Z.~Dubnickova, M.~A.~Ivanov, J.~G.~K\"orner, 
P.~Santorelli and G.~G.~Saidullaeva,
  %``One-photon decay of the tetraquark state 
  % $X(3872) \to \gamma + J/\psi$ in a relativistic constituent quark model 
  % with infrared confinement,''
  Phys.\ Rev.\ D {\bf 84}, 014006 (2011). 
%  doi:10.1103/PhysRevD.84.014006
%  [arXiv:1104.3974 [hep-ph]].

%\cite{Gutsche:2016cml} 
\bibitem{Gutsche:2016cml} 
  T.~Gutsche, M.~A.~Ivanov, J.~G.~K\"orner and V.~E.~Lyubovitskij,
  %``Isospin-violating strong decays of scalar single-heavy tetraquarks,''
  Phys.\ Rev.\ D {\bf 94}, 094012 (2016).
% arXiv:1608.00415 [hep-ph].

%\cite{Goerke:2016hxf}
\bibitem{Goerke:2016hxf} 
  F.~Goerke, T.~Gutsche, M.~A.~Ivanov, J.~G.~K\"orner, V.~E.~Lyubovitskij 
  and P.~Santorelli,
  %``Four-quark structure of Zc(3900), Z(4430) and Xb(5568) states,''
  Phys.\ Rev.\ D {\bf 94}, 094017 (2016). 
%  arXiv:1608.04656 [hep-ph].

%\cite{Ivanov:2005fd}
\bibitem{Ivanov:2005fd} 
  M.~A.~Ivanov, J.~G.~K\"orner and P.~Santorelli,
  %``Semileptonic decays of $B_c$ mesons into charmonium states 
  % in a relativistic quark model,''
  Phys.\ Rev.\ D {\bf 71}, 094006 (2005)
  [Phys.\ Rev.\ D {\bf 75}, 019901(E) (2007)]. 

%\cite{Efimov:1988yd}
\bibitem{Efimov:1988yd} 
  G.~V.~Efimov and M.~A.~Ivanov,
  %``Confinement and Quark Structure of Light Hadrons,''
  Int.\ J.\ Mod.\ Phys.\ A {\bf 4}, 2031 (1989); 
  I.~V.~Anikin, M.~A.~Ivanov, N.~B.~Kulimanova and V.~E.~Lyubovitskij,
  %``The Extended Nambu-Jona-Lasinio model with separable interaction:          
  %Low-energy pion physics and pion nucleon form-factor,''                      
  Z.\ Phys.\ C {\bf 65}, 681 (1995);
%\bibitem{Ivanov:1996pz}                                                        
  M.~A.~Ivanov, M.~P.~Locher and V.~E.~Lyubovitskij,
  %``Electromagnetic form factors of nucleons in a relativistic three-quark     
  %model,''                                                                     
  Few Body Syst.\  {\bf 21}, 131 (1996);
%  [arXiv:hep-ph/9602372];                                                      
  M.~A.~Ivanov, V.~E.~Lyubovitskij, J.~G.~K\"orner and P.~Kroll,
  %``Heavy baryon transitions in                                                
  %a relativistic three-quark model,''                                          
  Phys.\ Rev.\ D {\bf 56}, 348 (1997);
%  [arXiv:hep-ph/9612463];                                                      
%\bibitem{Ivanov:1999bk}                                                        
  M.~A.~Ivanov, J.~G.~K\"orner, V.~E.~Lyubovitskij and A.~G.~Rusetsky,
  %``Strong and radiative decays of heavy flavored baryons,''                   
  Phys.\ Rev.\ D {\bf 60}, 094002 (1999);
%  [arXiv:hep-ph/9904421];                                                      
%\bibitem{Faessler:2003yf}                                                      
  A.~Faessler, T.~Gutsche, M.~A.~Ivanov, V.~E.~Lyubovitskij and P.~Wang,
  %``Pion and sigma meson properties in a relativistic quark model,''           
  Phys.\ Rev.\  D {\bf 68}, 014011 (2003);
%  [arXiv:hep-ph/0304031];                                                      
%\bibitem{Ivanov:2006ni} 
  M.~A.~Ivanov, J.~G.~K\"orner and P.~Santorelli,
  %``Exclusive semileptonic and nonleptonic decays of the $B_c$ meson,''
  Phys.\ Rev.\ D {\bf 73}, 054024 (2006);
%  [hep-ph/0602050].
%\bibitem{Faessler:2006ft}                                                      
  A.~Faessler, T.~Gutsche, M.~A.~Ivanov, J.~G.~K\"orner,
  V.~E.~Lyubovitskij, D.~Nicmorus and K.~Pumsa-ard,
  %``Magnetic moments of heavy baryons in the relativistic                      
  %three-quark model,''                                                         
  Phys.\ Rev.\ D {\bf 73}, 094013 (2006);
%  [arXiv:hep-ph/0602193];                                                      
%\bibitem{Faessler:2006ky}                                                      
  A.~Faessler, T.~Gutsche, B.~R.~Holstein, V.~E.~Lyubovitskij,
  D.~Nicmorus and K.~Pumsa-ard,
  %``Light baryon magnetic moments and N --> Delta gamma                        
  %transition in a Lorentz covariant chiral quark approach,''                   
  Phys.\ Rev.\ D {\bf 74}, 074010 (2006);
%  [arXiv:hep-ph/0608015];                                                      
%\bibitem{Faessler:2008ix}                                                      
  A.~Faessler, T.~Gutsche, B.~R.~Holstein, M.~A.~Ivanov, J.~G.~K\"orner and
  V.~E.~Lyubovitskij,
  %``Semileptonic decays of the light J(P)=1/2(+) ground state baryon octet,''  
  Phys.\ Rev.\  D {\bf 78}, 094005 (2008);
%  [arXiv:0809.4159 [hep-ph]];                                                  
%\bibitem{Faessler:2009xn}                                                      
  A.~Faessler, T.~Gutsche, M.~A.~Ivanov, J.~G.~K\"orner and V.~E.~Lyubovitskij,
  %``Semileptonic decays of double heavy baryons in a relativistic constituent  
  %three-quark model,''                                                         
  Phys.\ Rev.\  D {\bf 80}, 034025 (2009);
%\bibitem{Branz:2009cd} 
  T.~Branz, A.~Faessler, T.~Gutsche, M.~A.~Ivanov, J.~G.~K\"orner 
  and V.~E.~Lyubovitskij,  
  %``Relativistic constituent quark model with infrared confinement,''
  Phys.\ Rev.\ D {\bf 81}, 034010 (2010); 
%  [arXiv:0912.3710 [hep-ph]].
%\bibitem{Ivanov:2011aa} 
  M.~A.~Ivanov, J.~G.~K\"orner, S.~G.~Kovalenko, P.~Santorelli and G.~G.~Saidullaeva,
  %``Form factors for semileptonic, nonleptonic and rare $B\, (B_s)$ meson decays,''
  Phys.\ Rev.\ D {\bf 85}, 034004 (2012);
%  [arXiv:1112.3536 [hep-ph]].
%\bibitem{Gutsche:2012ze}                                                       
  T.~Gutsche, M.~A.~Ivanov, J.~G.~K\"orner, 
  V.~E.~Lyubovitskij and P.~Santorelli,
  %``Light baryons and their electromagnetic interactions 
  %in the covariant constituent quark model,''                                                          
  Phys.\ Rev.\ D {\bf 86}, 074013 (2012);
%\bibitem{Gutsche:2013pp}                                                       
%  T.~Gutsche, M.~A.~Ivanov, J.~G.~K\"orner,
%  V.~E.~Lyubovitskij and P.~Santorelli,
  %``Rare baryon decays $\Lambda_b \to \Lambda {l^{+}l^{-}}                     
  %(l=e, \mu, \tau)$ and $\Lambda_b \to \Lambda\gamma$ :                        
  %differential and total rates, lepton- and hadron-side 
  %forward-backward asymmetries,''                                                                       
  Phys.\ Rev.\ D {\bf 87}, 074031 (2013); 
%\bibitem{Gutsche:2013oea}                                                      
%  T.~Gutsche, M.~A.~Ivanov, J.~G.~K\"orner, 
%  V.~E.~Lyubovitskij and P.~Santorelli,
  %``Polarization effects in the cascade decay 
  %$\Lambda_b \to \Lambda (\to p\pi\-) + J/\psi                                                                     
  %(\to \ell^+ \ell^-)$ in the covariant confined quark model,''                
  Phys.\ Rev.\ D {\bf 88}, 114018 (2013);
%\bibitem{Gutsche:2014zna}                                                      
%  T.~Gutsche, M.~A.~Ivanov, J.~G.~K\"orner, 
%  V.~E.~Lyubovitskij and P.~Santorelli,
  %``Heavy-to-light semileptonic decays                                         
  %of $\Lambda_b$ and $\Lambda_c$ baryons in the covariant confined quark model,''                                                                             
  Phys.\ Rev.\ D {\bf 90}, 114033 (2014);
%\bibitem{Gutsche:2015lea}                                                      
%  T.~Gutsche, M.~A.~Ivanov, J.~G.~K\"orner,
%  V.~E.~Lyubovitskij and P.~Santorelli,
  %``Towards an assessment of the ATLAS data on the branching ratio             
  %$\Gamma(\Lambda_b^0 \to \psi(2S)\Lambda^0)/\Gamma(\Lambda_b^0 \to J/\psi\Lambda^0)$,''                                                                      
  Phys.\ Rev.\ D {\bf 92}, 114008 (2015);
%\bibitem{Gutsche:2015rrt}                                                      
%  T.~Gutsche, M.~A.~Ivanov, J.~G.~K\"orner, 
%  V.~E.~Lyubovitskij and P.~Santorelli,
  %``Semileptonic decays $\Lambda_c^+ \to \Lambda \ell^+ \nu_\ell\,\,(\ell=e,\m\u)$ in the covariant                                                            
  %quark model and comparison with the new absolute branching fraction measurements of Belle and BESIII,''                                                     
  Phys.\ Rev.\ D {\bf 93}, 034008 (2016);
%\bibitem{Gutsche:2015mxa}                                                      
  T.~Gutsche, M.~A.~Ivanov, J.~G.~K\"orner, V.~E.~Lyubovitskij, P.~Santorelli 
  and N.~Habyl,
  %``Semileptonic decay $\Lambda_b \to \Lambda_c + \tau^- + \bar{\nu_\tau}$     
  %in the covariant confined quark model,''                                     
  Phys.\ Rev.\ D {\bf 91}, 074001 (2015)
  [Phys.\ Rev.\ D {\bf 91}, 119907(E) (2015)];
%\bibitem{Gutsche:2017wag}                                                      
  T.~Gutsche, M.~A.~Ivanov, J.~G.~K\"orner, V.~E.~Lyubovitskij,
  V.~V.~Lyubushkin and P.~Santorelli,
  %``Theoretical description of the decays                                      
  %$\Lambda_b \to \Lambda^{(\ast)}(\frac12^\pm,\frac32^\pm) + J/\psi$,''        
  arXiv:1705.07299 [hep-ph].

%\cite{Olive:2016xmw}                                                           
\bibitem{Olive:2016xmw}
  C.~Patrignani {\it et al.} [Particle Data Group],
  %``Review of Particle Physics,''                                              
  Chin.\ Phys.\ C {\bf 40}, 100001 (2016).

\end{thebibliography}
\end{document}